\documentclass[11pt,twoside]{article}
\usepackage{natbib,epsfig,verbatim,float,amsmath}
\usepackage{newpasp}
\def\A{{\bf A~}}
\def\B{{\bf B~}}
\def\An{{\bf A}}
\def\Bn{{\bf B}}
\def\edcomment#1{\iffalse\marginpar{\raggedright\sl#1\/}\else\relax\fi}
\reversemarginpar
\begin{document}
\title{Probing Relativistic Winds: The case of PSRJ07370-3039 \A \& \B}
 \author{Jonathan Arons and D.C. Backer}
\affil{University of California, Berkeley, Department of Astronomy, 601
Campbell Hall, Berkeley, CA 94720-3411}
 \author{Anatoly Spitkovsky}
\affil{Stanford University, Kavli Institute of Particle Astrophysics and 
Cosmology, P.O.Box 20540, MS 29, Stanford, CA 94309}
\author{V.M. Kaspi}
\affil{McGill University, Department of Physics, Montreal, QC H3A 2T8, Canada}

\begin{abstract}

We propose synchrotron absorption in a magnetosheath
forming a cocoon around the magnetosphere of pulsar \B to be the origin of the eclipse phenomena 
seen in the recently discovered double pulsar system 
PSRJ07370-3039 \A \& \Bn. The magnetosheath
enfolds the magnetosphere of pulsar \Bn, where the relativistic wind from
\A collides with \Bn's magentic field. If this model
is correct, it predicts the eclipses will clear at frequencies higher than those of the
observations reported to date (nominally, above $\nu \sim 5$ GHz.) The model also predicts 
synchrotron emission at the level of a few to 10 $\mu$Jy, peaking at $\nu \sim 2-5$ GHz
with possible orbital modulation. We use simplified semi-analytic models to elucidate
the structure of the \B magnetosphere, showing that the \A wind's dynamic pressure
confines \Bn's magnetic field to within a radius less than 50,000 km from \Bn, smaller than
\Bn's light cylinder radius, on the ``daytime'' side (the side facing \An).  Downstream
of \B (``nightime''), \B forms a magnetotail.  We use particle-in-cell simulations to
include the effects of magnetospheric rotation, showing that the magnetosheath 
has an asymmetric density distribution which may be
responsible for the observed eclipse asymmetries. We use simple estimates based
upon the magnetic reconnection observed in the simulations to derive a ``propellor'' 
spindown torque on \Bn, which is the dominant mode of angular mementum
extraction from this star.  Application of this torque to \Bn's observed spindown 
yields a polar dipole field  $\sim 7 \times 10^{11}$ Gauss (magnetic moment
$\mu_B \sim 3.5 \times 10^{29}$ cgs). This torque has a braking index of unity. 
We show that the model can explain the
known eclipses only if the \A wind's density is at least 4 orders of magnitude greater
than is expected from existing popular models of pair creation in pulsars.  We discuss
the implications of this result for our general understanding of pulsar physics.

Our proposal was qualitatively outlined in Kaspi {\it et al.} (2004) and
Demorest {\it et al.} (2004).  Since those papers' appearance, a similar proposal
has been made by Lyutikov (2004).

\end{abstract}

\section{Introduction \label{sec:introduction}}

Rotation Powered Pulsars (RPPs) lose their rotational energy
becuase of electromagnetic torques. While this fact has been known since the 
earliest days of pulsar research (Gold 1968, Goldreich and Julian 1969,
Ostriker and Gunn 1969), and indeed was predicted
before pulsars' discovery (Pacini 1967), 1) the physics of the
processes through which the extraction works, 2) the physics of how the 
rotational energy is transmitted to the surounding world, and 3) the
physics of how that energy transforms into the observed synchrotron 
radiation from the nebulae around pulsars have all
remained open questions.
Answers to all three questions are of significance not only
to the understanding of RPPs themselves, but also to the physics of
Active Galactic Nuclei and to the workings of Gamma Ray Burst sources, especially
if these outflows are driven by large scale Poynting fluxes from 
systematically magnetized disks (or perhaps magnetars, in the GRB case.)

Modern pulsar theory suggests
that a RPP throws off its rotational energy in the form of a relatively
dense, magnetized relativistic plasma wind, largely composed of
electron-positron pairs with an embedded wound up magnetic field.
Particle acceleration in electrostatic ``gaps'' (polar cap gaps, outer gaps or 
slot gaps) is thought to be the origin
of the $e^\pm$ plasma, through emission and conversion of gamma rays from 
accelerated particles within a RPP's magnetosphere ({\it e.g.} Hibschman and Arons 2001, 
Harding and Muslimov 1998, Muslimov and Harding 2003, Hirotani {\it et al.} 2003.)
The outflow densities suggested by 
these models justify the use of relativistic MHD in modeling the 
winds ({\it e.g.} Beskin, Kuznetsova and Rafikov 1998, Bogovalov 1999,
Contopoulos, Kazanas and Fendt 1999, Vlahakis 2004.) 
Theoretical models of MHD winds
exhibit negligible radiative emission (by construction), and indeed, there has been
no positive observational identification of the winds themselves - 
observational study of the winds' properties has depended on detection of the winds' 
consequences.  The winds are 
like a river flowing on dark nights -
invisible until the water strikes a dam, or rocks in the stream, when the glimmer
of starlight from the spray thrown by the obstacles allows one to infer the river's 
presence and properties.

To date, the main useful probe of RPPs' energy flow 
has been the winds' collisions with the ``dams'' created by
interstellar and circumstellar media surrounding RPPs. These collisions
create prominent Pulsar Wind Nebulae (PWNe) around the young pulsars 
with large rates of rotational energy loss $\dot{E}_R = c \Phi^2$
($\Phi$ is the electric potential drop across the magnetopsheric open field lines.) 
The radiative emissions from these nebulae
allow inferences
of the plasma content and magnetization at the winds' 
termination working surfaces (shock waves, in most interpretations.) See
Arons (1998, 2002, 2004), Slane (2002), Chevalier (2002, 2004), Reynolds (2003), Komissarov and
Lyubarsky (2003), Spitkovsky and Arons (2004), Del Zanna {\it et al.} (2004) 
for recent reviews and results on this class of interactions.

Rocks in the relativistic stream provide another window into relativistic wind
behavior. Examples of such interactions are
the collision between the wind and the ``excretion'' disk around the Be star
in PSR 1263-59 (e.g., Kaspi {\it et al.} 1995, Johnston {\it et al.} 1996, 2001, 
Tavani and Arons 1997), and the collision of the wind from the millisecond pulsar
PSR 1957+20 with the non-relativistic wind from its companion
star (Fruchter {\it et al.} 1988, Ruderman {\it et al.} 1989, Arons and Tavani 1993.)
As with the PWNe, most of what has been gleaned about the wind properties has come
from interpretations of the X-ray detections.

The recent discovery of the double pulsar system PSR J07370-3039 \A \& \B
(Burgay {\it et al.} 2003, Lyne {\it et al.} 2004) offers a new window into
studying a relativistic wind, in this case through the tools of radio astronomy.
The binary has an orbital period  $P_b$ of 2.4 hours, an orbital eccentricity 
$e = 0.08$ and an inclination angle $87 \pm 3$ degrees.
Pulsar  \A has a spin period
$P_{A}$ of 22.7 ms, a rotational energy loss rate  
$\dot{E}_{A}$ of  $0.6 \times 10^{34} \; {\rm erg~s}^{-1}$, 
and a light cylinder radius  
$R_{\rm LA}  =cP_A /2\pi = $1,084 km, which is small compared to the orbit
semi-major axis $a = (4.25 \mp 0.05) \times 10^{5}$ km.
Pulsar \B has a pulse period $P_B$ of 2.77 s, spin down rate 
$\dot{P}_B$ of  $0.8 \times 10^{-15}$ s s$^{-1}$ which leads to a rotational energy loss rate
$\dot{E}_B$ of  $2 \times 10^{30}$ ergs s$^{-1}$, and a light cylinder radius $R_{\rm LB}= 
1.32 \times 10^{5}$ km. Pulsar \A shows a brief eclipse that
lasts approximately 30 seconds when \A passes behind \B along the line 
of sight, at {\it inferior
conjunction}\footnote{This is standard terminology assuming that
\A is the {\it primary} star and \B the {\it secondary}. Likely binary 
evolution scenarios suggest
that this is the appropriate nomenclature for the two evolved remnants of the 
original main sequence system.
{\it Superior conjunction} occurs a half orbit later when \B passes behind 
\A along the sight line.}.
This eclipse shows substantially slower ingress (7 s) than egress (4 s), 
and the full eclipse profile is nearly achromatic
(Kaspi {\it et al.} 2004). The flux density and emission profile of pulsar \B vary around
the orbit in nearly achromatic manner over the range 430 MHz to 3.2 GHz (Lyne {\it et al.} 2004,
Demorest {\it et al.} 2004, Ramachandran {\it et al.} 2004). The
strongest emission episodes of \B are during  two orbital
longitude ranges about $70^\circ$ apart and asymmetrically spaced by $\sim30^\circ$ with
respect to inferior conjunction. Two weak 
\B emission episodes are  located
$\sim 115^\circ$ before inferior conjunction  (lasting $\sim40^\circ$) and 
$90^\circ$ after inferior conjunction  (lasting $\sim60^\circ$).  Pulsar \B is not
detected, or perhaps is seen with pulsed flux at the level of $ \sim 0.4 $\% of its maximum flux, 
during a range of orbital longitude that starts $\sim 60^\circ$ before
superior conjunction, and ends $\sim 30^\circ$ after this epoch - effectively, this
episode is an eclipse of \Bn.

These eclipses and emission episodes of \B offer an opportunity to probe the wind
around \A much closer to the energizing pulsar than has been possible
using PWNe in higher voltage systems.  Furthermore, the radio observations are
sensitive to low energy relativistic electrons and positrons, providing a look into
the instantaneous state of this component of a relativistic wind's plasma - PWNe observations
only constrain their winds' low energy particle content averaged over the
lives of the nebulae. 

We propose that \Bn's 
magnetosphere has a structure more similar to that of the Earth's
magnetosphere than to the magnetospheres of pulsars not interacting with a companion.
In contrast, \An's magnetospheric properties are decoupled from the binary.
Thus, the collision of \An's wind with \Bn's magnetosphere 
causes the formation of a bow shock. The pressure of the post-shock
particles and fields confines the \B magnetopshere on the side that
instantaneously faces \An, with a magnetotail extending behind \Bn.  Magnetic reconnection
allows the shocked wind to create a tangential stress on \Bn's magnetopshere, which
creates the dominant spin-down torque on \B (a variant of the propellor effect). There is
also a less significant relativistic wind component to the torque on \Bn, created by wind
from \B flowing out the magnetotail, whose transverse size is comparable to  $R_{\rm LB}$.
We speculate that the propellor torque also includes components that align \Bn's rotation axis
with the orbital angular momentum. If so, \B must be an orthogonal rotator, with its
magnetic axis perpendicular to its spin axis. 

Polarization observations obtained by Demorest {\it et al.} (2004) show that \A is almost
an aligned rotator (angle bewteen \An's magnetic and rotation axes $\sim 5^\circ$),
with its spin axis substantially misaligned with the orbital angular
momentum, $\angle ({\boldsymbol \Omega}_A , {\boldsymbol \Omega}_{\rm orbit}) \sim 50^\circ $.  
Then the equatorial current sheet in \An's wind is likely to have thickness 
$ \sim 10^\circ $ around the wind's equator, with the result that for $340^\circ$ of orbital phase,
\B is immersed in the high latitude, possibly slow and dense $e^\pm$ wind. We suggest
that latitudinal variation in the confining pressure exerted by the wind causes
variation of \Bn's beaming morphology, which may be in part responsible for the orbit dependent
variations of \Bn's pulse morphology.

The bow shock creates
a magnetosheath of relativistically hot, magnetized plasma  which enfolds 
the confined \B magnetosphere.  We show that synchrotron absorption in
the magnetosheath can explain the eclipse of \A at inferior conjunction and of \B
at superior conjunction. The model, which requires surprisingly high density in the
\A wind, predicts eclipse clearing at frequencies higher than
observed to date - nominally, above 5 GHz - and also predicts observable, orbitally
modulated synchrotron emission, at the level of 10 $\mu$Jy at $\nu \sim 5$ GHz.

In the context of this model the eclipses of  \A and \B and the other emission 
phenomenology of  \B provide
the first significant constraint on the properties of a relativistic wind
near its source  - \B forms a magnetospheric
rock in the relativistic stream from \A only 785 light cylinder radii outside the fast
pulsar's magnetosphere. The model suggests that $\sigma$, the ratio of 
Poynting flux to kinetic energy flux, in 
\An's wind  just upstream of the bow shock is certainly less than 2.5 and probably 
$\sim 0.2$, much less than what is expected from existing ideal MHD 
theories of relativistic wind outflow. Thus, the interaction of \An's wind with \B 
suggests magnetic dissipation in the wind begins quite close to the source.

\section{Magnetospheric Shape and Torques \label{sec:shape}}

We follow geophysical 
practice and identify the direction from \B to \A as ``noon'' as seen
from \Bn, with ``daytime'' being the hemisphere facing \An. The simulations  
described below show that the rotating magnetsphere develops polar cusps
in the magnetic field as each pole rotates past noon, features which correspond to
rotationally induced asymmetry of the magnetosheath. Using this asymmetry to
interpret the asymmetry of the 
\A eclipse suggests the \B pulsar has prograde rotation
with respect to the orbit, which identifies ``dawn'' 
and ``dusk'' as the directions
parallel and antiparallel to B's orbital velocity, respectively. 
``Midnight'' is in the downstream direction, looking down the magnetotail.
Balancing the full relativistic dynamic pressure of \An's wind against \Bn's
magnetic pressure yields a force balance radius at the magnetospheric apex 
on the daytime side
\begin{equation}
R_{m0}  =  \left( \frac{8 \mu_B^2 a^2 c}
         { \dot{E}_A \sin^2 \Theta_{AB}} \right)^{1/6} 
\simeq  4.84 \times 10^4 \left(\frac{\mu_{B,30}}{0.375}\csc \Theta_{AB} \right)^{1/3} 
           \; {\rm km}, \label{eq:mpause_nose} 
\end{equation}
where $\mu_B $ is \Bn's magnetic moment, $\mu_{B,30} = \mu_B /10^{30}$ cgs and
$\Theta_{AB} $ is the angle between \An's rotation axis and \Bn's orbital
position - \An's wind has dynamic pressure varying with latitude.  Since
$\sin^2 \Theta_{AB} = 1 - \sin^2 i_A \sin^2 (\omega - \psi_A)$,
the confining dynamic pressure varies by about a factor of two, twice during
each orbit. Therefore 
the size of \Bn's daytime magnetosphere and its polar cap vary with orbital longitude $\omega$.
Here $i_A $ is the angle between \An's angular velocity ${\boldsymbol \Omega_A}$
and the orbital angular momentum, probably $\sim 50^\circ $ (Demorest {\it et al.} 2004), 
and $\psi_A$
is the angle between the projection of ${\boldsymbol \Omega_A}$ on the orbital
plane and the line of apsides at $\omega = 0$. 

\begin{figure}[H]
\unitlength = 0.0011\textwidth
\begin{center}
(a)
\begin{picture}(300,300)(0,15)
\put(0,0){\makebox(300,300){ \epsfxsize=300\unitlength \epsfysize=300\unitlength
\epsffile{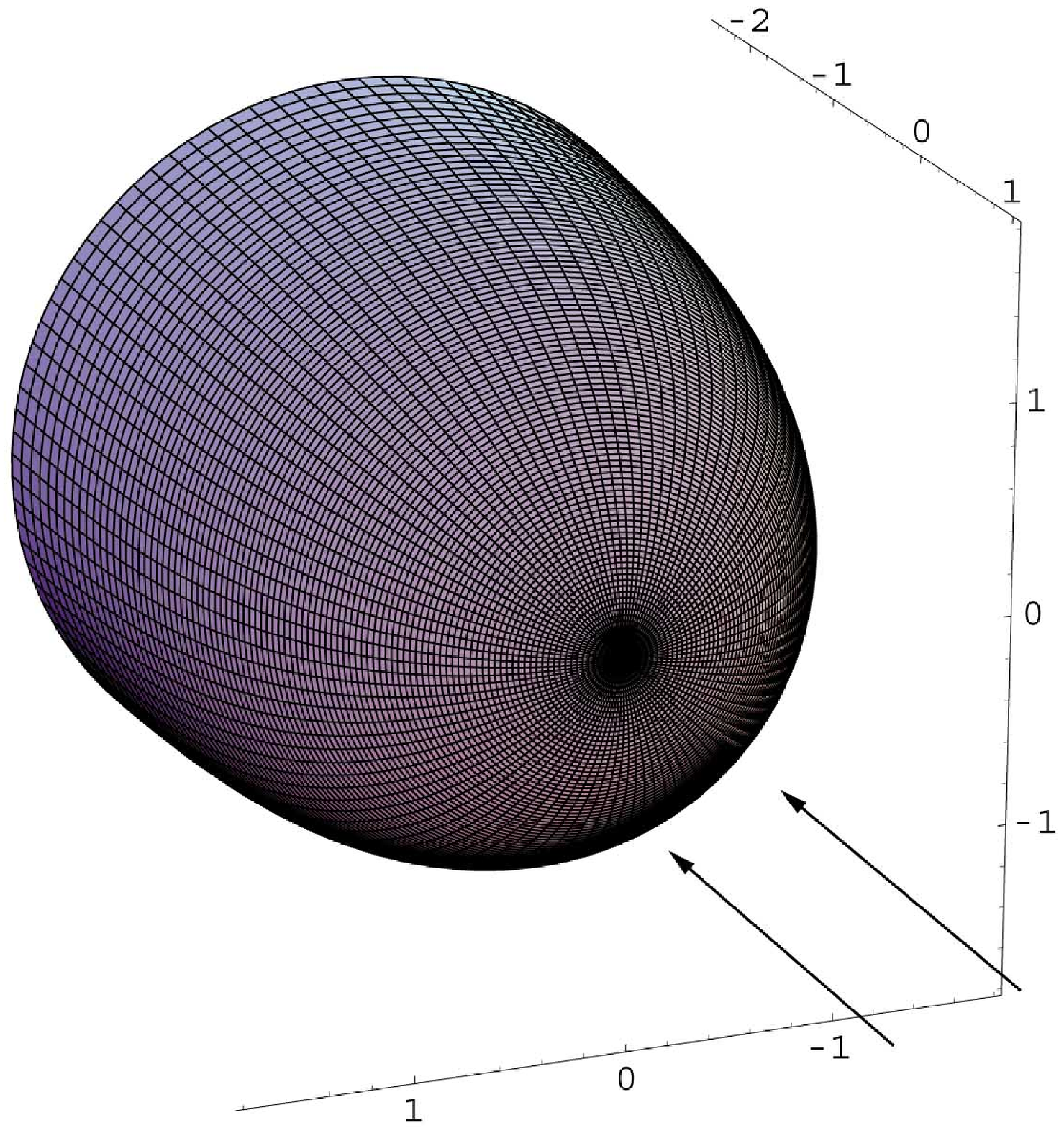}
}}
\end{picture}
\hspace{50\unitlength}
(b)
\begin{picture}(300,300)(0,15)
\put(0,0){\makebox(300,300){ \epsfxsize=300\unitlength \epsfysize=300\unitlength
\epsffile{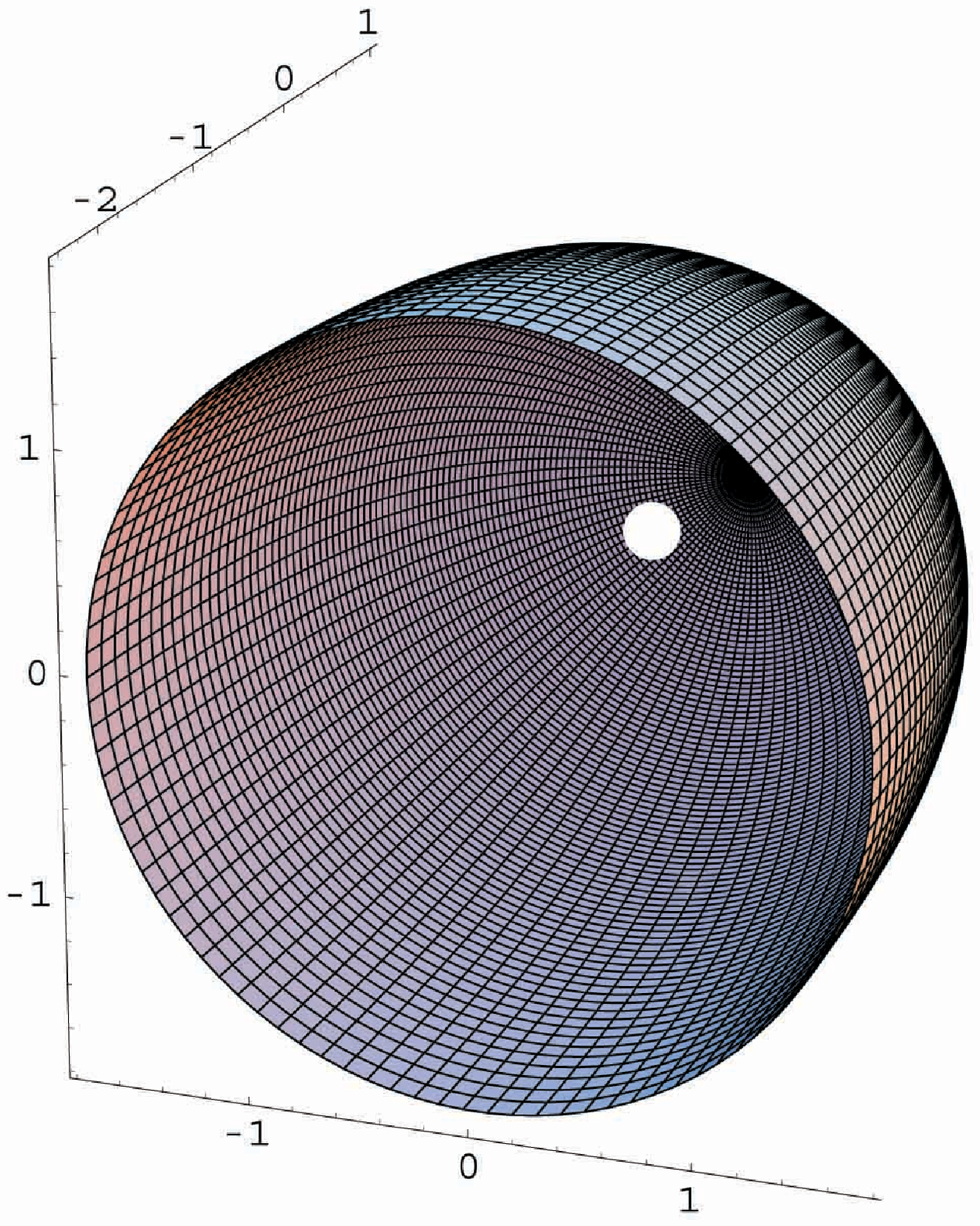}
}}
\end{picture}
\hspace{50\unitlength}\\
(c)
\begin{picture}(300,300)(0,15)
\put(0,0){\makebox(300,300){ \epsfxsize=300\unitlength \epsfysize=300\unitlength
\epsffile{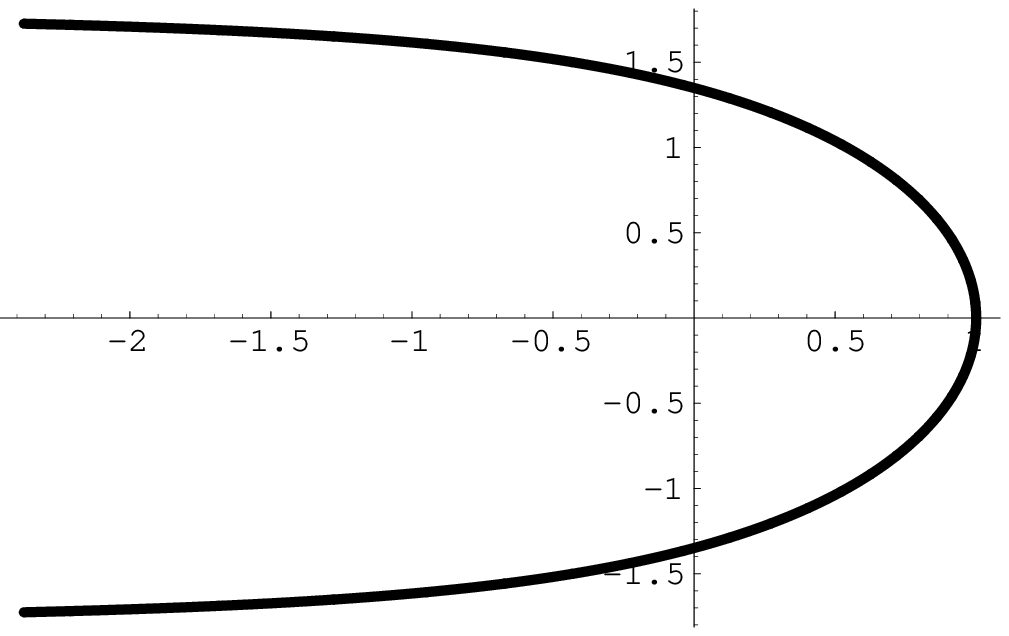}
}}
\end{picture}
\end{center}
\caption{Shape of B's magnetosphere, in a steady flow, axisymmetric pressure
equilibrium model. (a) The magnetospheric obstacle as seen from upstream in A's wind. 
(b) The magnetospheric obstacle as seen looking up the magnetotail. The B neutron star is the
white dot (not to scale). (c) A cross section of the magnetopause surface, taken
through the axis of symmetry.  The magnetopause radius is measured in units of $R_{m0}$.
\label{fig:shape} }
\end{figure}

Figure \ref{fig:shape} shows the
shape of a non-rotating \B magnetopshere obtained from the pressure balance condition.

\begin{figure}[H]
\unitlength = 0.0011\textwidth
\begin{center}
(a)
\begin{picture}(300,250)(0,15)
\put(-10,0){\makebox(300,250){ \epsfxsize=300\unitlength \epsfysize=300\unitlength
\epsffile{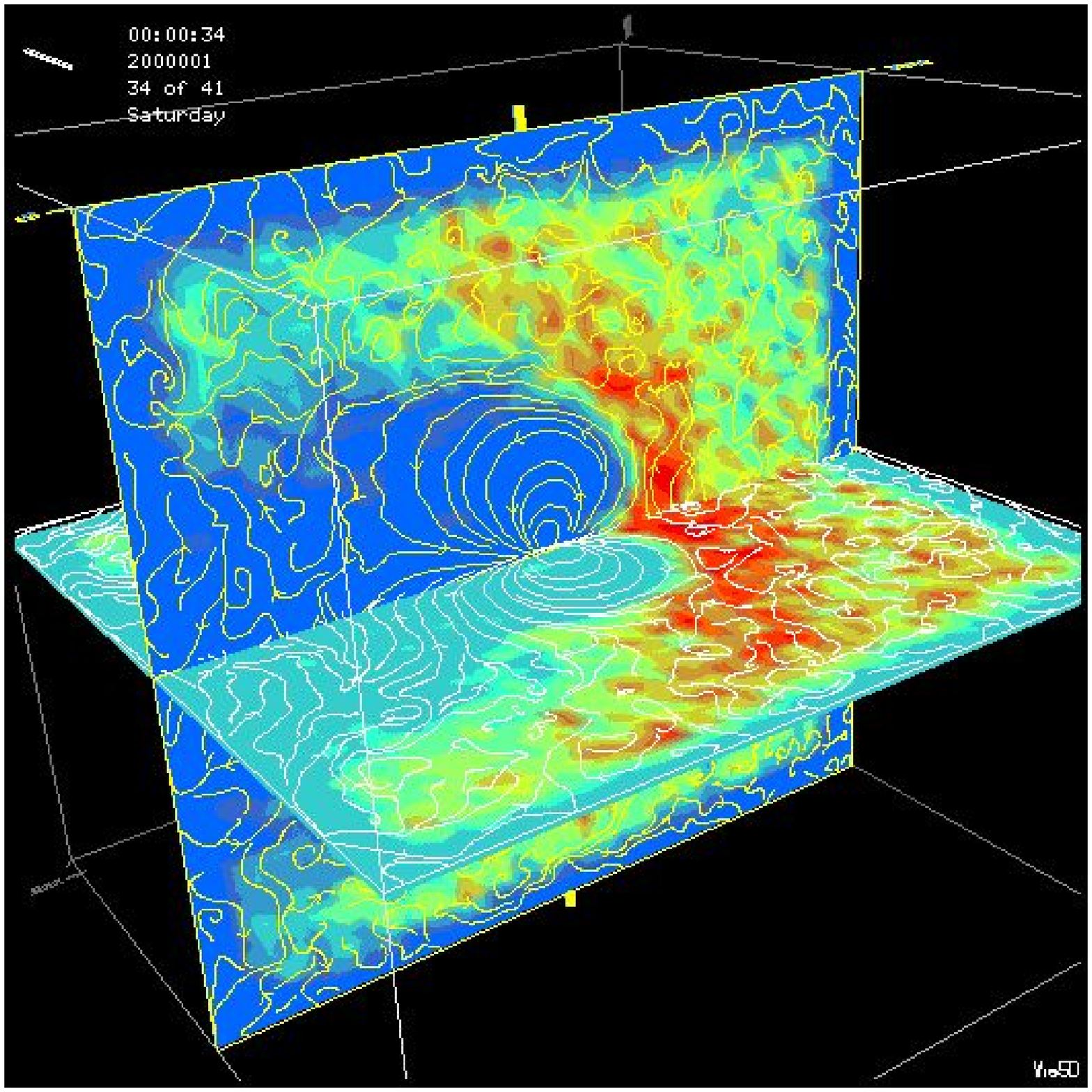}
}}
\end{picture}
\hspace{50\unitlength}
(b)
\begin{picture}(400,250)(0,15)
\put(0,0){\makebox(400,250){ \epsfxsize=400\unitlength \epsfysize=300\unitlength
\epsffile{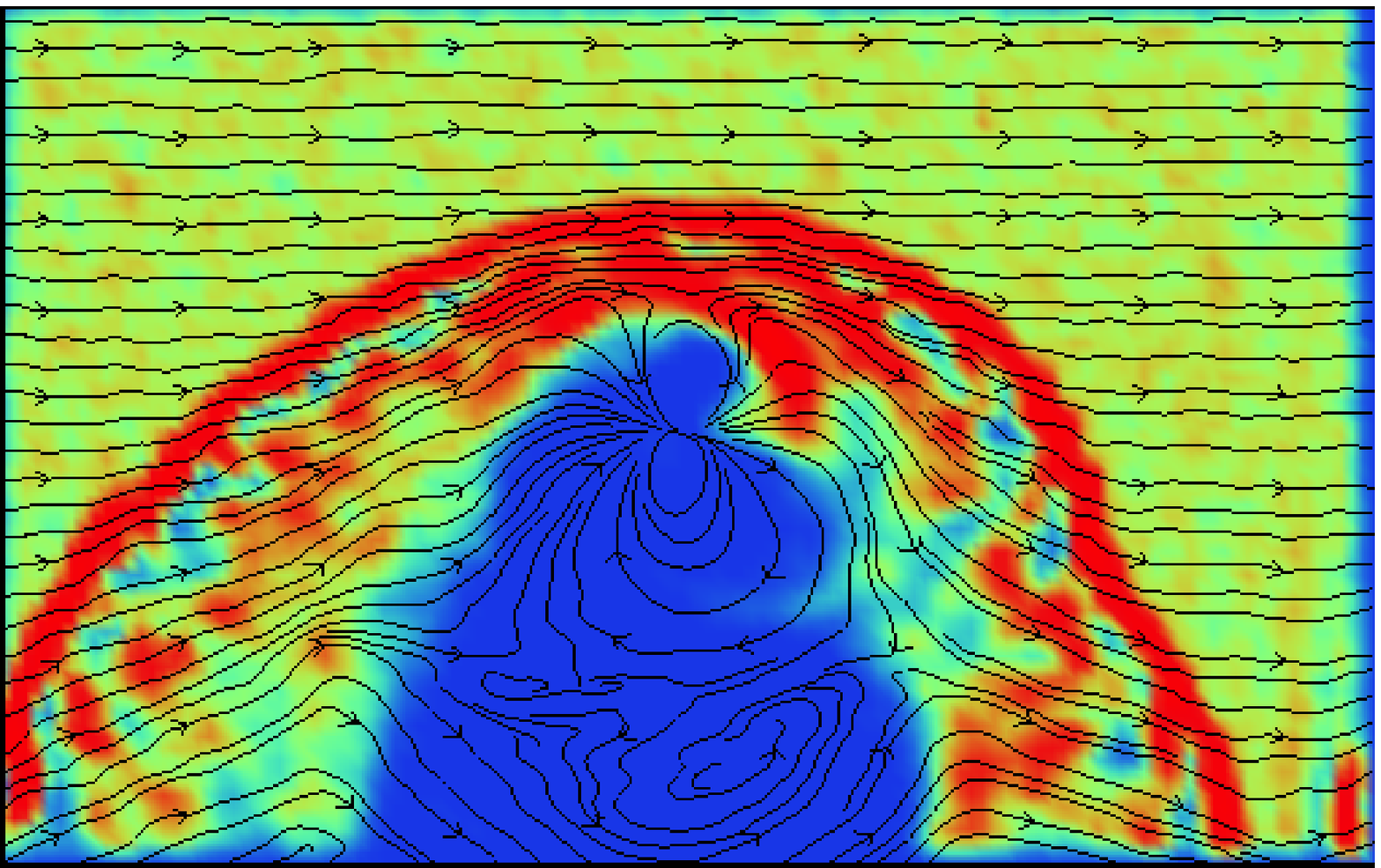}
}}
\end{picture}
\end{center}
\caption{(a) Relativistic 3D PIC simulation of \Bn's rotating magnetosphere immersed in an
unmagnetized wind. The magnetospheric obstacle 
creates a shock heated magnetosheath on the daytime side - the wind approaches from the 
right.  Darker shading indicates the higher density regions in the magnetosheath.
Partial trapping 
of plasma in the cusp creates an asymmetric density structure, with excess
plasma in the magnetosheath's afternoon sector. 
(b) Relativistic 3D PIC simulation of \Bn's rotating magnetosphere immersed in a
{\it magnetized} wind - a snapshot of the equatorial plane. The magnetic field structure shown 
is in the half period of \Bn's rotation
when the wind's and \Bn's magnetic fields are oppositely
directed.  Reconnection causes field lines to cross the magnetospheric boundary,
creating tangential drag on \Bn's magnetsophere, which leads to the
dominant torque on \Bn.   Note the
high density region over the polar cusp at ``3 o'clock.'' In the alternate half period, 
the upstream magnetosphere is closed,
with magnetospheric structure similar to the equatorial plane of (a).
\label{fig:sim}}
\end{figure}

We have carried out a series of particle-in-cell (PIC) simulations of the \B
magnetosphere's structure, with \Bn's rotation included. Figure \ref{fig:sim}
shows snapshots, with \B assumed to
be an orthogonal rotator.

One can readily show that the tangential stress exerted
by the reconnected field on the rotating magnetosphere of \B exerts a torque on \B which is 
larger than the relativistic torque that would be present if \B were isolated, essentially
because the magnetopause radius $R_m$ is small compared to $R_{\rm LB} $.  This model 
(equivalent to a propellor effect torque, but with a physical basis in field dynamics well
known from geophysical magnetopsheres) yields the torque
\begin{eqnarray}
(\dot{J}_B)_{rec} & \approx &  -\frac{1}{3} \frac{{\dot E}_A \sin^2 \Theta_{AB}}{a^2 c} 
        \frac{\Omega_B R_{m0}}{c} R_{m0}^3 \nonumber \\
& \approx &  \left(\frac{{\dot E}_A \sin^2 \Theta_{AB}}{a^2 c} \right)^{1/3}
    \mu_B^{4/3} \frac{\Omega_B}{c} \label{eq:rec-torque} \\
& = & 3.3 \times 10^{30} \mu_{B,30}^{4/3} \sin^{2/3} \Theta_{AB} \; {\rm ergs}.
	\nonumber
\end{eqnarray}

Using (\ref{eq:rec-torque}) and writing $\dot{E}_B = -\Omega_B \dot{J}_B $ yields 
$\mu_B = 0.375 \times 10^{30}$ cgs ($B_{pole} = 2 \mu_B /R_*^3 = 0.75 \times 10^{12}$ Gauss),
about a factor of 3 less than the value of \Bn's magnetic moment found by
Lyne {\it et al.} (2004), who used the standard relativistic torque for an isolated RPP
to estimate the equatorial surface field of \Bn. The torque due to a relativistic wind
from \B (which must flow out the tail, the only region with enough room
to allow \Bn's rotation to wind up the magnetic field) 
is less than 25\% of the torque due to dayside reconnection.

\section{Synchrotron Absorption in the Magnetosheath}

Existing relativistic shock theory (Kennel and Coronoti 1984a, Lyutikov 2004)
allows us to evaluate the properties of the
shocked magnetosheath plasma at the nose of the magnetosphere, where the shock and the
magnetic field are transverse to the flow.   The upstream pair density
in the wind is 
$n_{1\pm} = \mu_A \kappa /4P_A ec R_{LA} a^2 = 0.023 \kappa$ cm$^{-3}$,
while the upstream magnetic field is $B_1 = \mu_A /2 R_{LA}^2 a = 6.3 $ Gauss, with
corresponding cyclotron frequency $\nu_{g1} = 18$ MHz. Currently popular pair creation 
models suggest $\kappa \sim 10-100$ when applied to \An.

The eclipse profile exhibited by pulsar A requires the obscuring plasma to 
form a belt lying in the plane formed by the line of sight as it passes over (under) pulsar
\B with impact parameter $b$. The 30 second eclipse duration around superior conjunction
corresponds to a belt length 18,600 km, oriented in the
direction of the pulsars' relative motion (Kaspi {\it et al.} 2004).
The light curve of the 
B pulsar shows a period of $\sim 90^\circ$ in orbital phase centered somewhat before
inferior conjunction when B almost disappears (Ramachandran {\it et al.} 2004.) 

We suggest these eclipses are natural consequences of the absorbtion in the magnetosheath
enfolding pulsar B - at superior conjunction, the observer looks up the optically thin
magnetotail through the absorbing magnetosheath toward \An, as shown in Figure \ref{fig:shape}(b), 
while for a wide range of orbital phase around
inferior conjunction, the magnetosheath absorbs \Bn's pulsed radiation along the line of sight
to the observer, as is apparent from Figure \ref{fig:shape}(a). This interpretation
implies the absorption to be confined to an apex cap on \Bn's magnetopause,
which extends from the magnetospheric axis out to a coltatitude of
$ \theta_{ec} \sim 45^\circ $ from that axis. The simulations shown in Figure \ref{fig:sim}
are roughly consistent with an absorbing cap of this size. Elementary geometry applied 
to the solution shown in Figure \ref{fig:shape}(c), with the length of \An's eclipse specified to
be 18,600 km, then yields $b = 47,900 $ km = 0.16 lt-sec, for magnetospheric scale
$R_{m0} = 48,400$ km (expression \ref{eq:mpause_nose}) and shock standoff distance
at the beginning and end of the eclipse
$\Delta_s \approx R_m(\theta_{ec}) \beta_2$.  This impact parameter corresponds to
our view of the system being $3.2^\circ$ off the orbital plane, or $i = 86.8^\circ$, a value
consistent with the measured $i = 87^\circ \pm 3^\circ$.

The asymmetry in the plasma density apparent in Figure \ref{fig:sim} is a candidate to
explain the asymmetry in the \A eclipse only if \Bn's rotation is prograde with the
orbital motion.

We assume the bow shock converts flow energy
into isotropic non-thermal particle distributions with the
power law form in given by $N_{2\pm} (\gamma) = (s-1) n_{2\pm} \gamma^{-s}$, with 
$s > 2$ and $\gamma \geq \gamma_m $. The MHD shock jump conditions yield the
post shock temperature $T_2$,  density $n_{2\pm}$ and velocity $\beta_2$, all functions
of $\sigma = B^2 /4\pi m_\pm n_{1\pm} \gamma_{wind} $ just upstream of the shock. 
The post-shock temperatue specifies  $\gamma_m = (s-2) T_2 /(s-1)$. We assume
$\gamma_{wind} = \sigma_0 /(1 + \sigma)$, where 
$\sigma_0 = (B^2 /4 \pi m_\pm n_\pm)_{r=R_{\rm LA}} = 7.5 \times 10^7 /\kappa $.

\begin{figure}[H]
\unitlength = 0.0011\textwidth
\begin{center}
(a)
\begin{picture}(250,250)(0,15)
\put(0,0){\makebox(250,250){ \epsfxsize=290\unitlength \epsfysize=290\unitlength
\epsffile{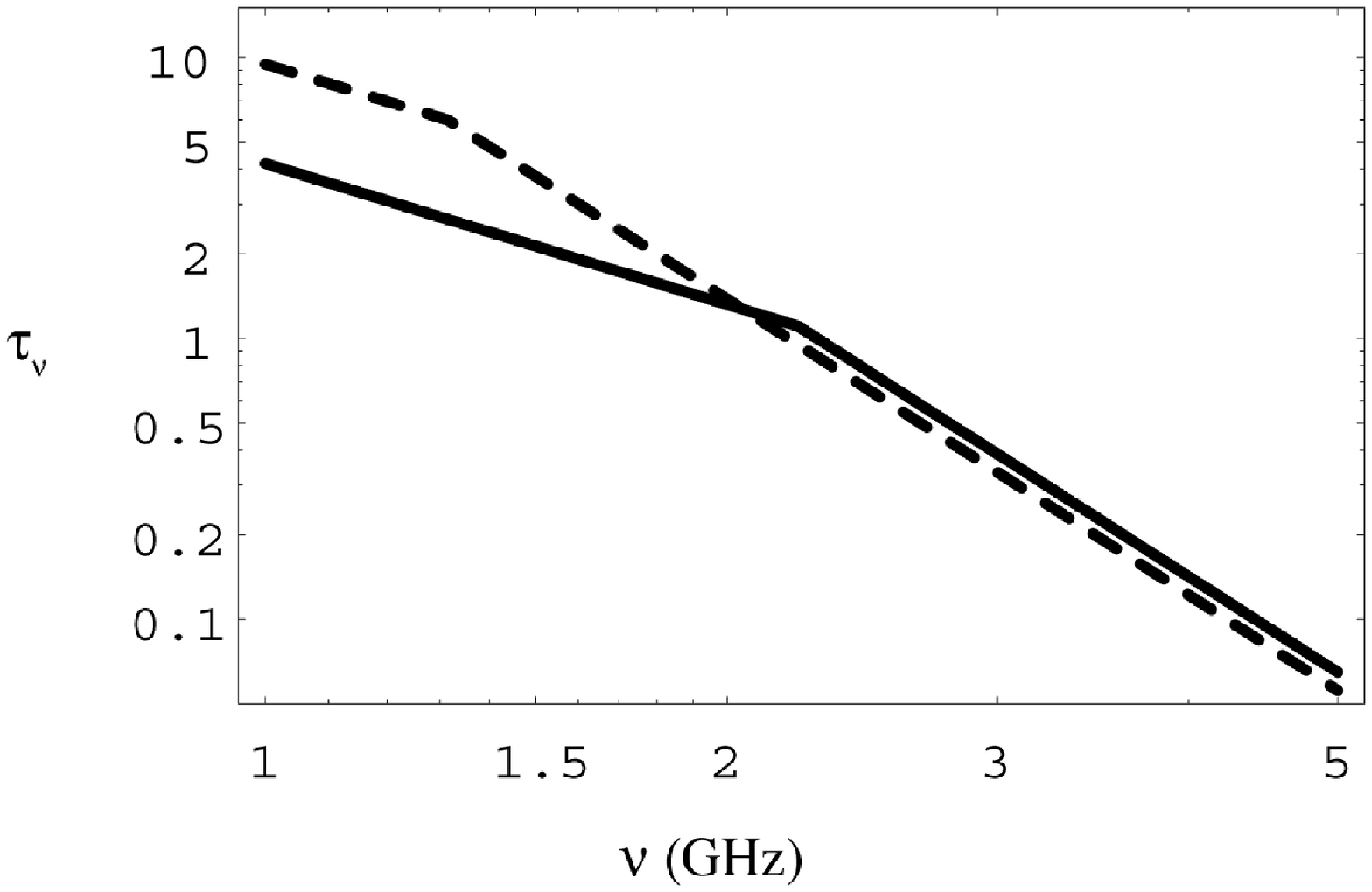}
}}
\end{picture}
\hspace{50\unitlength}
(b)
\begin{picture}(250,250)(0,15)
\put(0,0){\makebox(250,250){ \epsfxsize=290\unitlength \epsfysize=290\unitlength
\epsffile{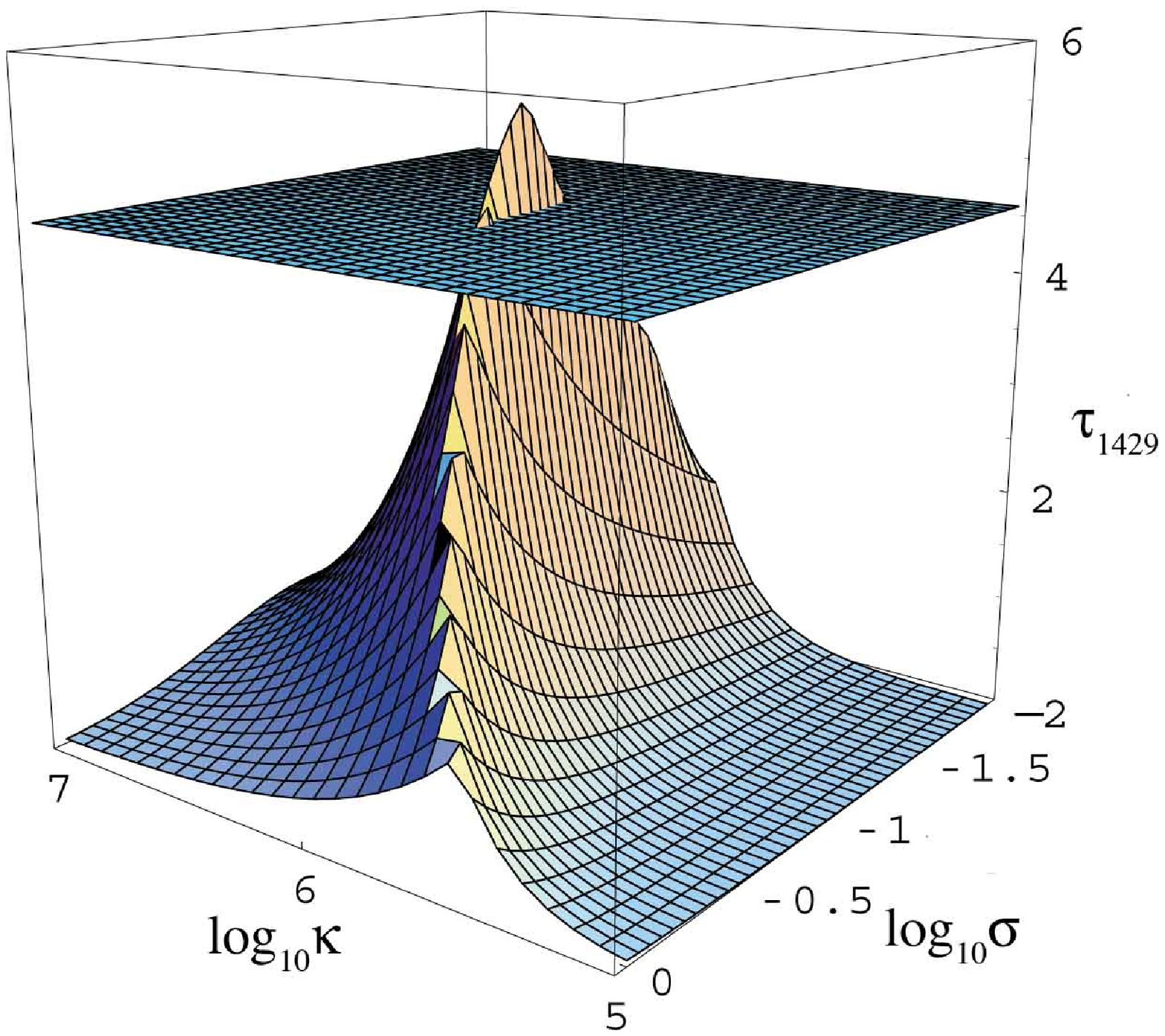}
}}
\end{picture}
\hspace{50\unitlength} \\
(c)
\begin{picture}(250,250)(0,15)
\put(0,0){\makebox(250,250){ \epsfxsize=290\unitlength \epsfysize=290\unitlength
\epsffile{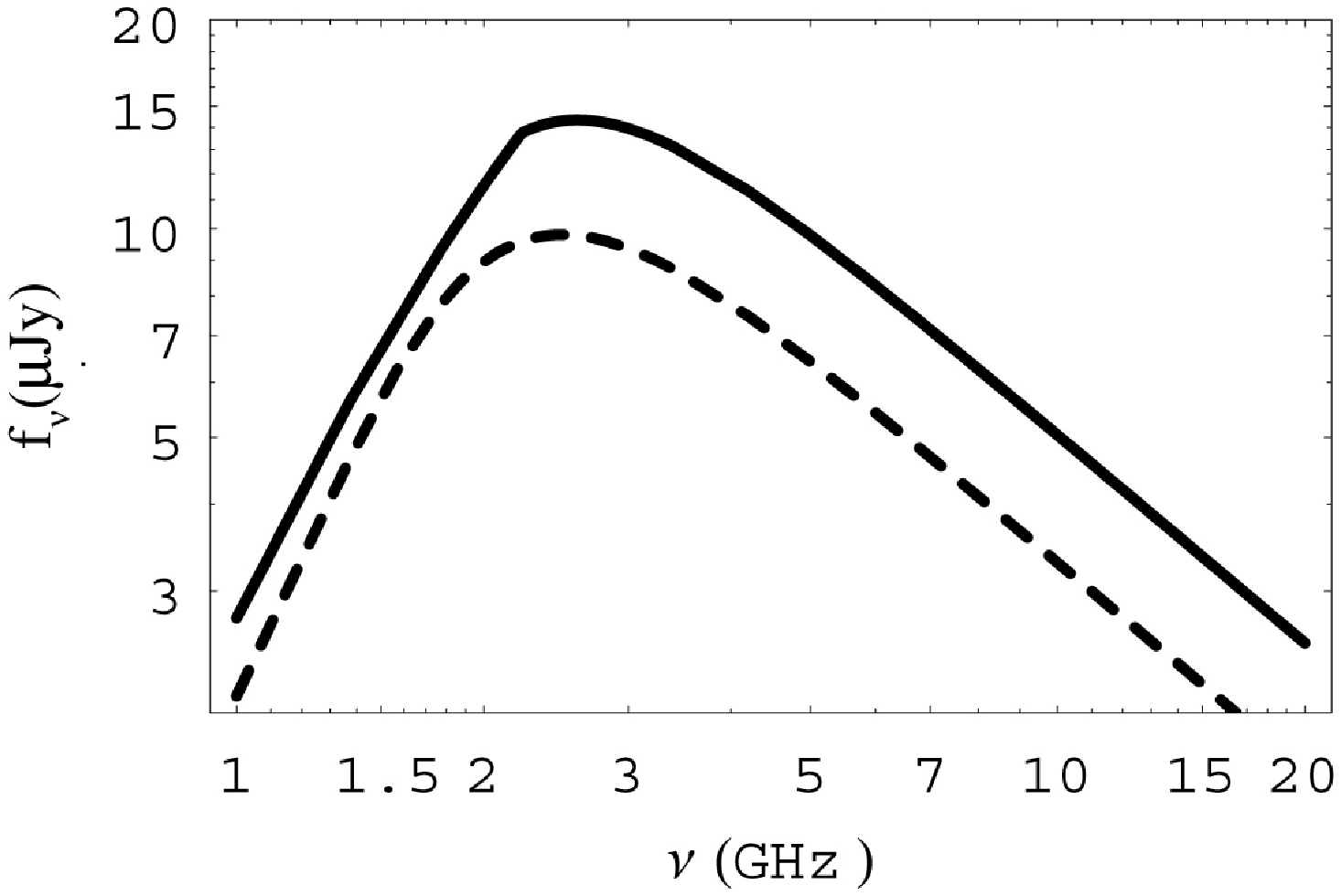}
}}
\end{picture}
\hspace{50\unitlength}
(d)
\begin{picture}(250,250)(0,15)
\put(0,0){\makebox(250,250){ \epsfxsize=290\unitlength \epsfysize=290\unitlength
\epsffile{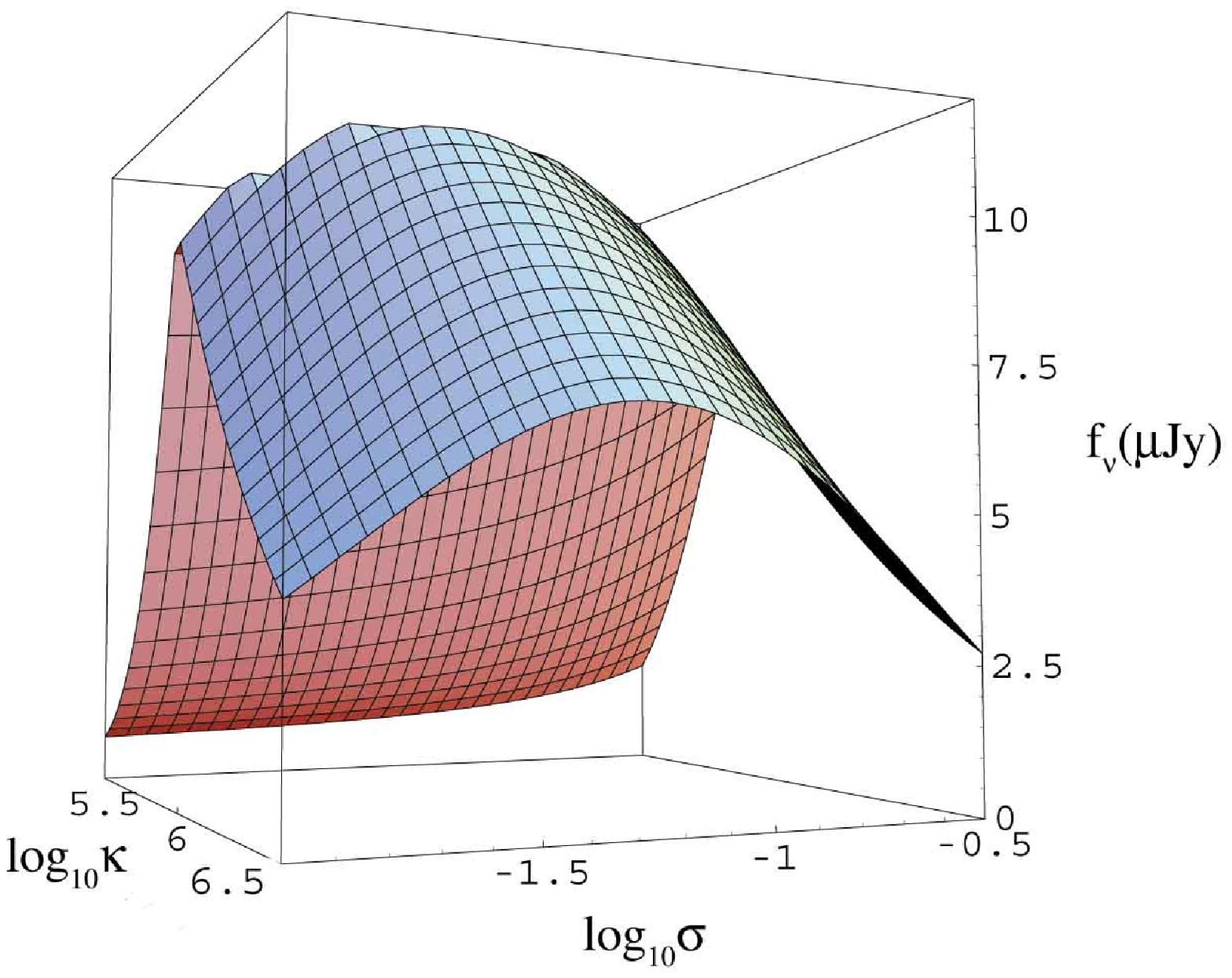}
}}
\end{picture}
\end{center}
\caption{(a)  Non-thermal optical depth through the apex magnetosheath as a function 
of $\nu$ for the models 
with maximum non-thermal magnetosheath emission (solid curve), 
$\kappa = 10^{6}, \; {\rm and} \; \sigma = 0.03$;
minimum density consistent with the eclipse of pulsar A at 1429 MHz (long-dashed curve),
$\kappa = 10^{6}, \; {\rm and} \; \sigma = 0.2$.
(b) Non-thermal optical depth at 1429 MHz as a 
function of wind density parameter $\kappa$, and of wind $\sigma$ just upstream of the
bowshock. The level plane corresponds to optical depth 4.6 (flux at eclipse center
1\% of the unobscured flux). (c) Non-thermal synchrotron 
spectra of the magnetosheath as a function of $\nu$ for the models 
with maximum magnetosheath emission (solid curve), 
$\kappa = 10^{6}, \; {\rm and} \; \sigma = 0.03$;
minimum density consistent with the eclipse of pulsar A at 1429 MHz (long-dashed curve),
$\kappa = 10^{6}, \; {\rm and} \; \sigma = 0.2$.
(d) Non-thermal synchrotron spectra at 2 GHz as a 
function of density parameter $\kappa$ and of wind $\sigma$. These figures assume 
a nonthermal particle distribution with $s=3$ (optically thin emission $\propto \nu^{-1}$),
the distance to the
pulsars to be 500 pc, a nose radius $R_{m0} = 48,400$ km, and the emitting region to occupy a
cap on the apex of the magnetopshere with opening angle $45^\circ$.}
\label{fig:non-thermal_model}
\end{figure}

Standard results for the synchrotron opacity yield the optical depth 
through the magnetosheath at the apex, where we take the magnetosheath thickness
to be $R_{m0} \beta_2 $, scaling found in the terrestrial bow shock (Spreiter {\it et al.} 1966)
and also found in our relativistic simulations.
Figures \ref{fig:non-thermal_model}(a) and (b), constructed for the case $s = 3$, show that
optical depth at 1429 MHz adequate to explain the eclipses of \A and \B
requires high density in \An's wind, $\kappa \approx 10^6 $, and low wind magnetization,
$\sigma \approx 0.1 $. 
\noindent Non-thermal 
emission from the magnetosheath might be detectable. 
At high frequencies the magnetosheath becomes optically thin, with emission spectrum varying 
in proportion to $\nu^{-(s-1)/2 }$; at low frequency $f_\nu \propto \nu^{5/2}$. These
frequency dependencies neglect the local inhomogeneity of the magnetosheath obvious in 
Figure {\ref{fig:sim}(b). Figure \ref{fig:non-thermal_model}(c) shows the 
emission spectra for models
corresponding to enough optical depth to explain the \A eclipse, and to yield the 
maximum emission from the magnetosheath, while  Figure \ref{fig:non-thermal_model}(d) 
shows the dependence of the emission flux at 5 GHz on upstream density and on $\sigma$.

\section{Discussion and Conclusions}

Our results show that if synchrotron absorption in the magnetosheath is the cause
of eclipse phenomena in this fascinating system, the wind from \A at latitudes outside
the equatorial current sheet ($|\lambda | > 5^\circ $) is dense ($\kappa \sim 10^6$), slow
($\gamma_{wind} \sim 75$) and weakly magnetized at $r \approx 850,000 $ km from \An.
The magnetosheath synchrotron absorption model predicts the eclipses will clear at
higher frequencies (nominally, $\nu > 5$ GHz), and that synchrotron emission (probably
with some orbital modulation) will be detectable at the level of 5-10 $\mu$Jy at
$\nu \sim 5$ GHz.

If this is the correct interpretation of the eclipse phenomena, the eclipses are the
first (semi-)direct detection of a RPP's wind outside of the equatorial current sheet.
Essentially all the phenomena in the young PWNe can be ascribed to the equatorial winds
(Kennel and Coroniti 1984a,b, Coroniti 1990, Gallant and Arons 1994, Bogovalov and Khangoulian
2002, Lyubarsky 2002, Komissarov and Lyubarsky 2003, 2004, Spitkovsky and
Arons 2004, Del Zanna {\it et al.} 2004) as they interact with the plasma of the surrounding
PWNe. The density paramater $\kappa$ is very high compared to expectations derived from
current models of magnetospheric pair creation. These have been reasonably successful in
accounting for the plasma fluxes inferred to be in the equatorial winds, where we see the results
of a very high $\gamma_{wind} $ outflow.  They have not been successful in accounting for
the larger populations of lower energy particles which produce the radio synhrotron
emission from the young PWNe.  The low energy particle injection rates averaged over
the history of these systems are factors of 50,  and more, larger than derived
from standard pair creation models (polar cap, outer gap,  slot gap,...).  Our results,
and analogous results found by Lyutikov (2004), are even more radical - standard pair creation 
models applied to \A yield pair injection rates at least 4 orders of magnitude smaller
($\kappa < 100$) than are required by the magnetosheath synchrotron absorption model.

One solution to the excess low energy plasma problem in young PWNe has been unusual
evolutionary spindown history. If the energizing pulsars had much larger spin rates (or magnetic
fields) early in their lives than one derives from a constant braking index, constant magnetic
moment model, the pair production rates at earlier times might have been much larger,
and the equatorial winds much slower, than they are at present. This is a possible (if perhaps
unlikely) resolution of the problem, since radio emitting electrons and positrons in the PWNe
live ``forever'', with synchrotron lifetimes much in excess of the PWNe ages.

Such an evolutionary solution cannot explain the plasma overdensity inferred here for \An's
wind - plasma striking \Bn's magnetosphere emerged from \A only 3 seconds before it enters
the magnetosheath, and flows out of the magnetosheath even more quickly, after absorbing
pulsed radiation both from \A and from \Bn. The most efficient explanation of the discrepancy
is that the standard pair creation theories are inadequate (i.e., {\it wrong}), as applied
to field lines which feed the non-equatorial wind. We point out that where pair creation
models have worked reasonably well, they apply to the feeding of the equatorial wind, and to
the origin of pulsed gamma rays - both phenomena occur on field lines connected to the stars near
the boundary between the closed and open field lines of their magnetospheres.

Following Ruderman and Sutherland's (1975) vacuum gap model of polar cap pair creation,
all subsequent theories have assumed an electrostatic gap structure (strictly steady in the 
co-rotating frame), with pair plasma taking on the role of poisoning the gap accelerator
as fast as the density builds up. The spatial rate of such build up varies, depending on
gap geometry and dominant gamma ray emission and pair creation processes, but in all
cases, the production rates required by radio observations are not achieved.  In the
case of pulsar \A, one can readily show that if one gives up the concept of gap poisoning
and simply asssumes that particle acceleration along the magnetopsheric magnetic field
continues uninihibited by pair creation, as in Tademaru's (1973) early cascade model, 
pair outflows from \A as large as we have infered from the magnetosheath absorption model
are possible.  The physics behind such behavior remains to be elucidated.

The PSRJ07370-3039 \A \& \B system clearly has  promise for helping
us to unravel the mysteries of relativistic outflows from compact objects, as well as
providing fascinating phenomenological food for thought and further study.  The details
of the model described here, along with a number of aspects not touched on in this brief
report, will be reported elsewhere.

\end{document}